\documentclass[sigconf]{acmart}
\renewcommand\footnotetextcopyrightpermission[1]{}
\def\BibTeX{{\rm B\kern-.05em{\sc i\kern-.025em b}\kern-.08emT\kern-.1667em\lower.7ex\hbox{E}\kern-.125emX}}
\usepackage{hyperref}
\usepackage{nicefrac}
\usepackage{siunitx}
\usepackage{array,framed}
\usepackage{booktabs}
\usepackage{
  color,
  float,
  epsfig,
  wrapfig,
  graphics,
  graphicx,
  subcaption
}
\usepackage{textcomp}
\usepackage{setspace}
\usepackage{latexsym,fancyhdr,url}
\usepackage{enumerate}
\usepackage[noend]{algpseudocode}
\usepackage{graphics}
\usepackage{xparse} 
\usepackage{xspace}
\usepackage{multirow}
\usepackage{csvsimple}
\usepackage{balance}

\usepackage{
  tikz,
  pgfplots,
  pgfplotstable
}
\usepackage{bbm}
\usepackage{hyperref}
\usepackage{algorithm}
\usepackage{booktabs}
\usepackage{float}
\floatname{algorithm}{Protocol}
\usetikzlibrary{
  shapes.geometric,
  arrows,
  external,
  pgfplots.groupplots,
  matrix
}

\pgfplotsset{compat=1.9}

\usepackage{mathtools,}

\DeclareMathAlphabet{\mathcal}{OMS}{cmsy}{m}{n}

\DeclareGraphicsExtensions{%
    .png,.PNG,%
    .pdf,.PDF,%
    .jpg,.mps,.jpeg,.jbig2,.jb2,.JPG,.JPEG,.JBIG2,.JB2}

\usepackage{xparse}
\newcommand{\A}{\vec{\mathbf{a}}}
\newcommand{\B}{\vec{\mathbf{b}}}
\newcommand{\D}{\vec{\mathbf{d}}}
\renewcommand{\S}{\vec{\mathbf{s}}}
\newcommand{\T}{\vec{\mathbf{t}}}

\newcommand{\bnm}{\begin{newmath}}
\newcommand{\enm}{\end{newmath}}

\newcommand{\bea}{\begin{eqnarray*}}%
\newcommand{\eea}{\end{eqnarray*}}%

\newcommand{\bne}{\begin{newequation}}
\newcommand{\ene}{\end{newequation}}

\newcommand{\bal}{\begin{newalign}}
\newcommand{\eal}{\end{newalign}}

\newenvironment{newalign}{\begin{align}%
\setlength{\abovedisplayskip}{4pt}%
\setlength{\belowdisplayskip}{4pt}%
\setlength{\abovedisplayshortskip}{6pt}%
\setlength{\belowdisplayshortskip}{6pt} }{\end{align}}

\newenvironment{newmath}{\begin{displaymath}%
\setlength{\abovedisplayskip}{4pt}%
\setlength{\belowdisplayskip}{4pt}%
\setlength{\abovedisplayshortskip}{6pt}%
\setlength{\belowdisplayshortskip}{6pt} }{\end{displaymath}}

\newenvironment{newequation}{\begin{equation}%
\setlength{\abovedisplayskip}{4pt}%
\setlength{\belowdisplayskip}{4pt}%
\setlength{\abovedisplayshortskip}{6pt}%
\setlength{\belowdisplayshortskip}{6pt} }{\end{equation}}

\newcounter{ctr}

\newcounter{mytable}
\def\mytable{\begin{centering}\refstepcounter{mytable}}
\def\endmytable{\end{centering}}

\newcounter{myfig}
\def\myfig{\begin{centering}\refstepcounter{myfig}}
\def\endmyfig{\end{centering}}

\newlength{\saveparindent}
\newlength{\saveparskip}
\newcommand{\E}{{\rm I\kern-.3em E}}

\renewcommand{\eqref}[1]{\mbox{Equation~(\ref{#1})}}

\def \part {part}

\renewcommand{\paragraph}[1]{\vspace*{6pt}\noindent\textbf{#1}\;}

\def \blackslug{\hbox{\hskip 1pt \vrule width 4pt height 8pt
    depth 1.5pt \hskip 1pt}}
\def \qed{\quad\blackslug\lower 8.5pt\null\par}

\newcounter{mynote}[section]

\newcommand\ignore[1]{}

\newcounter{rcnote}[section]

\newcounter{mrnote}[section]

\newcounter{fknote}[section]

\newcounter{anote}[section]

\DeclareMathSymbol{\mlq}{\mathord}{operators}{``}
\DeclareMathSymbol{\mrq}{\mathord}{operators}{`'}

\newcommand{\rhf}[2]{R_{f, \gamma}}

\DeclareDocumentCommand{\edist}{o o}{
  \ensuremath{
    \IfNoValueTF{#1}{{d}}{{\sf d}(#1,#2)}
  }
}

\newcommand{\olrk}[1]{\ifx\nursymbol#1\else\!\!\mskip4.5mu plus 0.5mu\left(\mskip0.5mu plus0.5mu #1\mskip1.5mu plus0.5mu \right)\fi}

\NewDocumentCommand{\indseq}{ O{1} O{r} }{{#1}\ldots {#2}}

\begin{document}
\fancyhead{}

\def\thetitle{FIAT: Fine-grained Information Audit for Trustless Transborder Data Flow}
\title{\thetitle}

\author{Shuhao Zheng}
\email{shuhao.zheng@mail.mcgill.ca}
\affiliation{McGill University \& ZK0 Labs}
\author{Yanxi Lin}
\email{lin-yx18@tsinghua.org.cn}
\affiliation{Tsinghua University}
\author{Yang Yu}
\email{yangyu1@mail.tsinghua.edu.cn}
\affiliation{Tsinghua University}
\author{Ye Yuan}
\email{ye.yuan3@mail.mcgill.ca}
\affiliation{McGill University}
\author{Yongzheng Jia}
\email{abner@berkeley.edu}
\affiliation{UC Berkeley}
\author{Xue Liu}
\email{xueliu@cs.mcgill.ca}
\affiliation{McGill University}

\date{}

\begin{abstract}
  Auditing the information leakage of latent sensitive features during the transborder data flow has attracted sufficient attention from global digital regulators.
  However, there is missing a technical approach for the audit practice due to two technical challenges.
  Firstly, there is a lack of theory and tools for measuring the information of sensitive latent features in a dataset. 
  Secondly, the transborder data flow involves multi-stakeholders with diverse interests, which means the audit must be trustless.  
  Despite the tremendous efforts in protecting data privacy, an important issue that has long been neglected is that the transmitted data in data flows can leak other regulated information that is not explicitly contained in the data, leading to unaware information leakage risks.
  To unveil such risks trustfully before the actual data transfer, we propose \textbf{FIAT}, a \textbf{F}ine-grained \textbf{I}nformation \textbf{A}udit system for \textbf{T}rustless transborder data flow.
  In FIAT, we use a learning approach to quantify the amount of information leakage, while the technologies of zero-knowledge proof and smart contracts are applied to provide trustworthy and privacy-preserving auditing results.
  Experiments show that large information leakage can boost the predictability of uninvolved information using simple machine-learning models, revealing the importance of information auditing.
  Further performance benchmarking also validates the efficiency and scalability of the FIAT auditing system.

\end{abstract}



\maketitle
\section{Introduction}
The globalization of global digital economy relies on the free information flow over the Web infrastructures. 
In particular, the transborder data flows play an important role in the technology improvement and business activities of international digital companies and digital trades \cite{kunerRegulationTransborderData2011}.
However, the concern about privacy and national security drives the regulators to set requirements on the transborder data flows.
For example, many developing countries such as counties in African have adopted frameworks to regulate the transborder data flows \cite{mauritiusDataProtectAct2017}.
International organizations such as OECD \cite{OECDGuidelinesProtection}, European Union \cite{EUdataregulations2018}, and APEC \cite{cooperation2005apec} have also published specific guidelines on protecting data privacy for transborder data flows, which have been widely adopted and executed.

There have been great efforts in designing systems to implement these data protection regulations for data privacy \cite{sterling,castaldo2018blockchain,casalini2019trade,rahman2020accountable,liuDealerEndtoendModel2021,songZKDETTraceablePrivacyPreserving}.
With advanced tools from information theory, cryptography, and blockchain technology, these systems could achieve provable privacy protection of personal data in practical and efficient ways.
Moreover, designs like Sterling \cite{sterling} and Dealer \cite{liuDealerEndtoendModel2021} also support data receivers to require computation over the data from data senders and obtain the computation results, enlarging the application scope for more complex real-world scenarios.
As most of the data are eventually consumed by automatic data processing algorithms for value exploitation \cite{liangSurveyBigData2018} which usually require well-processed data representations, data representation learning \cite{bengio2013representation} takes a large portion of the computation demands in such systems.

However, the current relative works have not systematically addressed the challenge of latent-information leakage during transborder data flow. 
Many cases and studies have shown that the privacy and national security information can be leaked during the transborder data flow even if the data flow does not directly include those pieces of sensitive features.
The rivalry agent can infer the information about the sensitive features from the data by applying the current powerful data mining tools. 
For instance, it is possible to infer users' privacy profiles (e.g., age, gender) from their click and purchase records \cite{hasanDiscussionPrivacyChallenges2013}.
Therefore, the privacy information that uniquely identifies a person's racial, religious, or political status, as specified in Art 9.1 of the European General Data Protection Regulation (GDPR) \cite{EUdataregulations2018}, could all be leaked through the data flow.
The situation can get even worse if the transborder data flow is related to national security, e.g., the statistics of city transportation can implicitly contain positional information about military bases.
Moreover, if the scope of acceptable computation is not clearly specified, a malicious data receiver can exploit the sensitive information by carefully designing the computation.

Therefore, it is quite necessary to audit the risk of information leakage of latent features in transborder data flows \cite{koshiyamaAlgorithmAuditingSurvey2021}.
However, two obstacles challenge the audit on the transborder data flow. 
Firstly, there is missing a method to verify whether and to what extent the audited data set includes the information about the latent sensitive features.
Secondly, the transborder data audit involves various stakeholders with divergent interests.
Even if there exists an auditing method, the divergent interests can drive the auditor to temp the auditing results.
Further, the auditor can also leak the data intentionally.
Thus, it is a technical challenge how to design a proper auditing system that presents trustworthy auditing results while still preserving data security.

To address the above-mentioned challenges, in this paper, we propose \textbf{FIAT}, a \textbf{F}ine-grained
\textbf{I}nformation \textbf{A}udit system for \textbf{T}rustless transborder data flow.
Specifically, we propose to quantify the information leakage risks by measuring the mutual information between the regulated information and the data representations transmitted in the data flow.
We also visualize the auditing results through predictions with machine learning.
In this way, FIAT provides both an information-theoretic and practical estimation of the real information leakage risk.
Moreover, we adopt zero-knowledge proof (ZKP) \cite{goldwasserKnowledgeComplexityInteractive1985} and smart contracts in FIAT to enable trustworthy and privacy-preserving information auditing before data transfer.
ZKP guarantees the correctness of all the computations during auditing and the privacy of the data taken into auditing.
Meanwhile, the smart contract serves as a publicly verifiable record to validate the correctness of the zero-knowledge proof computation, thus enhancing the trustfulness of the auditing results.
Hence, FIAT enables fine-grained, trustless, and privacy-preserving information leakage auditing for transborder data flows. 

We conduct extensive experiments to unveil the aforementioned information leakage risks quantitatively on Adult Data Set \cite{Dua:2019}, a public dataset that contains personal information collected from adults all over the world.
To present the information leakage estimation more concretely, we train a simple machine learning model to predict the regulated information using the data or data representations that are previously considered safe for transfer.
Results show that data with high information leakage can greatly boost the prediction accuracy of the regulated information.
Furthermore, we implement the FIAT auditing system using \textbf{circom} \cite{Circom2022} and \textbf{snarkjs} \cite{Snarkjs2022}, and conduct benchmarking evaluations to characterize the practical performance of information auditing on real-world datasets.
Benchmarking results illustrate the feasibility of using ZKP for data auditing, i.e., a complete auditing circuit for a dataset with $5000$ pieces of personal information has only $<2^{24}$ R1CS constraints, which can be proved by cutting-edge ZKP protocols in minutes \cite{drevonBenchmarkingZeroKnowledgeProofs}.
Moreover, with Groth16 \cite{grothSizePairingBasedNoninteractive2016}, the verification of the proof only takes $2$ seconds and the proof size is only $0.8KB$, which makes it practical to be uploaded to the blockchain and verified in smart contracts.

To the best of our knowledge, this is the first work to present a theoretical model to quantify information leakage risks, and further implement both trustless and efficient information auditing systems using zero-knowledge proofs. We summarize the key contributions of this work as follows: 

\begin{itemize}
    \item We present the importance of information auditing for regulated information that is not explicitly presented in transborder data flows, which has long been neglected in previous data privacy research. To further illustrate our theory, we propose a computation model using mutual information and machine learning to visualize such information leakage risks.
    \item We adopt advanced zero-knowledge proof protocols and smart contracts to provide efficient, trustless and privacy-preserving information auditing for transborder data flows.
    \item We design and implement \textbf{FIAT}, a \textbf{F}ine-grained \textbf{I}nformation \textbf{A}uditing system that integrates the above techniques for \textbf{T}rustless transborder data flows.
    Evaluation results show the efficiency and scalability of the \textbf{FIAT} system in information auditing for real-world data flows.
\end{itemize}
\section{Preliminaries}
In this section, we provide the necessary background on mutual information, zero-knowledge proofs, and smart contract.
\subsection{Mutual Information \& Entropy}
C.E. Shannon~\cite{theoryOfCommuniation} built the foundation of the information theory, which first defined and analyzed the "quantity" between two random variables.
Such quantity was formulated as mutual information \cite{termMI}.
Mutual information is a statistical tool used to measure the dependency between two random variables, which has already been widely adopted in various areas like machine learning \cite{tishbyDeepLearningInformation2015} and data pricing \cite{liFirstLookInformation2017}.
Compared to the correlation coefficient, which is restricted to real-valued random variables and linear dependence, mutual information is more extensive in quantifying the amount of information, such as in bits or nats, obtained about one random variable while observing another random variable.
By leveraging the mutual information, we are able to assess how different the joint distribution of the pair $(X, Y)$ is from the product of the marginal distributions of $X$ and $Y$.
In information theory, the average level of intrinsic "information" of a random variable can be expressed as the entropy of this random variable.
The entropy of a given discrete random variable $X$ is defined as
\begin{equation}
    H(X) := \mathbb{E}[-\log p(X)] = -\sum_{x \in \mathcal{X}} p(x)\log p(x),
\end{equation}
where the domain of $X$ is denoted as $\mathcal{X}$, and $X$ is distributed according to $p:{\mathcal{X} \to [0,1]}$ such that $p(x):=\mathbb{P} [X=x]$.
One can further formulate the conditional entropy between two random variables X and Y from $\mathcal{X}$ and $\mathcal{Y}$ respectively as
\begin{equation}
    H(X|Y) := -\sum_{x,y \in \mathcal{X} \times \mathcal{Y}} p_{X, Y}(x, y)\log\frac{p_{X, Y}(x, y)}{p_{Y}(y)},
\end{equation}
where $p_{X, Y}(x, y) := \mathbb{P}[X=x, Y=y]$ and $p_{Y}(y) := \mathbb{P}[Y=y]$.
The conditional entropy represents the amount of randomness in the random variable $X$ that still exists when giving the random variable $Y$.
The mutual information between the random variables $X$ and $Y$ can be defined as
\begin{equation}
    I(X; Y) := H(X) - H(X|Y) \equiv H(Y) - H(Y|X),
\end{equation}
where the $H(X)$ and $H(Y)$ are the marginal entropy, and $H(X|Y)$ and $H(Y|X)$ are the conditional entropy.

\subsection{Zero-knowledge Proof \& zk-SNARK}
Zero-knowledge proof (ZKP) is a family of cryptographical protocols that enable a prover to prove to a verifier that she knows some secret without leaking any information about that secret, first proposed by Goldwasser et al. \cite{goldwasserKnowledgeComplexityInteractive1985}.
In practice, most ZKP applications adopt ZKP to prove $\mathcal{NP}$ relations, i.e., computationally hard problems whose answers are easy to verify, such as circuit satisfiability problems.
We show a formal definition of ZKP for $\mathcal{NP}$ relations in Appendix \ref{appendix:zkp_def}.
Most computations can be represented as circuits with inputs and outputs.
The performance of a ZKP system for circuit satisfiability is usually measured by the circuit size, i.e., the number of gates in the arithmetic circuit.
The efficiency metrics include the proof size, the prover's running time, and the verifier's running time.

zk-SNARK \cite{bitanskyExtractableCollisionResistance2012}, known as \textit{zero-knowledge succinct non-interactive arguments of knowledge}, is a set of recent ZKP protocols that achieves practical adoption in blockchain systems, such as Groth16 \cite{grothSizePairingBasedNoninteractive2016}, PLONK \cite{gabizonPLONKPermutationsLagrangebases2019b}, Virgo \cite{zhangTransparentPolynomialDelegation2020}, etc.
As the name shows, zk-SNARK algorithms are non-interactive, and usually have small proof size, short prover time and verifier time, making them suitable for real-world applications which usually have large circuits.
\subsection{Smart Contract}
A smart contract is a transparent computer program that is automatically executed on the blockchain \cite{savelyev2017contract}.
Each smart contract is an on-chain account itself.
The execution of a smart contract is automatically triggered by the transactions sent from other accounts, and the transactions must contain formatted data.
Ethereum \cite{buterin2014next} has become the most popular smart contract platform in terms of market cap and available DApps and activities \cite{schar2021decentralized}.
Other smart contract platforms include Solana \cite{yakovenko2018solana}, Polkadot \cite{wood2016polkadot}, Hyperledger Fabric \cite{androulaki2018hyperledger}, etc.

\begin{figure*}
    \centering
    \includegraphics[width=0.7\textwidth]{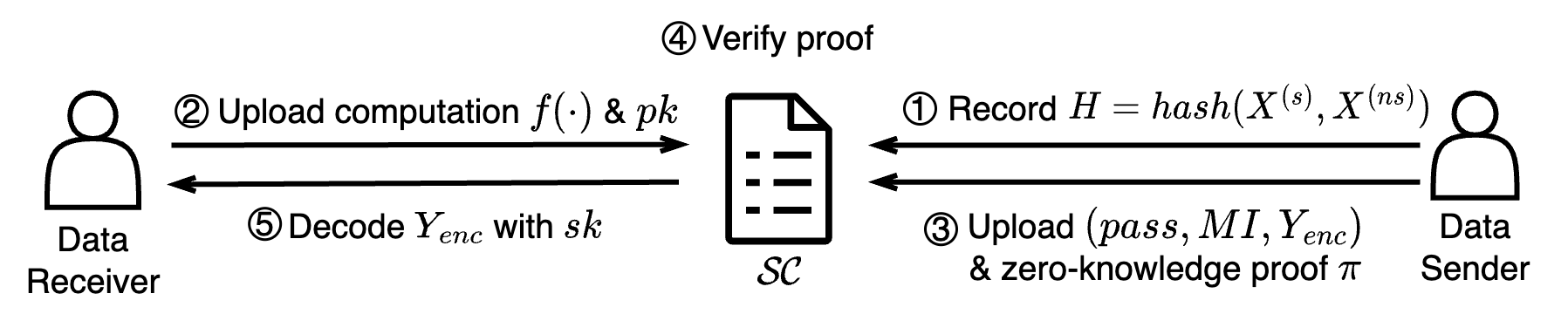}
    \caption{An Overview of the FIAT System}
    \label{fig:system}
\end{figure*}
\section{Related Work}
In this section, we introduce pioneering works on protecting data privacy for data flows and show the importance of information auditing.
Then, we briefly show some recent progress on representation learning, which is the main data computation considered in the FIAT system.

\subsection{Data Privacy}
In recent years, the ubiquitous transborder data transfer has become a hot topic that draws attention from policy-makers and researchers around the world.
In 1980, OECD issued Guidelines on the Protection of Privacy and Transborder Flows of Personal Data \cite{OECDGuidelinesProtection} to regulate transborder data flows of its members, which was further revised in 2013.
Moreover, GDPR \cite{EUdataregulations2018}, a regulation on data protection and privacy in the European Union and the European Economic Area, has become an important part of EU privacy law and human rights law since its adoption in 2016.
Besides these guidelines, there are also great efforts in designing specific regulation principles and systems for transborder data flows.
Kuner \cite{kunerRegulationTransborderData2011} summarized the status-quo of the transborder data flow regulation, providing a systematic view of regulations in the past and future.
Casalini \& González \cite{casalini2019trade} provided an overview of ways to regulate cross-border data flows.
Castaldo \& Cinque \cite{castaldo2018blockchain} proposed a blockchain-based system for health data exchange that guarantees traceability and liability.
Rahman et al. \cite{rahman2020accountable} designed an accountable cross-border data sharing platform that is able to handle the misbehaving of all participants in the platform.

A specific line of work for privacy-preserving data transfer focus on the data flows in data market.
Sterling \cite{sterling}, a decentralized marketplace for data computing, adopted Trust Execution Environments (TEE) to enable privacy-preserving analysis and machine learning over private data.
Liu et al. \cite{liuDealerEndtoendModel2021} designed Dealer, in which data valuation and privacy protection are both done through differential privacy.
zkDET \cite{songZKDETTraceablePrivacyPreserving} exploited the zero-knowledge proof and non-fungible token (NFT) to provide data traceability to track the origin and transformations of datasets, as well as protect data privacy while ensuring fairness during the data exchanging process.

Another line of work recently proposed to use zero-knowledge proof to protect the privacy of data owners during federated learning.
Zero-knowledge proof is a powerful tool to protect data privacy and guarantee system security at the same time.
Guo et al. \cite{guoSecureWeightedAggregation2021} and Burkhalter et al. \cite{burkhalterRoFLAttestableRobustness2022} adopted zero-knowledge proof to enhance the secure aggregation of local model weights.
Nguyen \& Thai \cite{nguyenPreservingPrivacySecurity2022} asked each data owner to submit a zero-knowledge proof for their local model training to defend against malicious data owners.
In light of these pioneering works, we also adopt zero-knowledge proof in designing the auditing system to protect information security in a new aspect.

\subsection{Representation Learning}
Data representation is a fundamental issue of modern machine learning algorithms that helps convert the information in the raw data into low-dimensional vectors.
Bengio et al. \cite{bengio2013representation} defined representation learning as learning representations of the data that make it easier to extract useful information when building classifiers or other predictors.
Nowadays, representation learning is widely applied in many different areas.
For instance, Mikolov et al. \cite{mikolov2013distributed} proposed a model for learning vector representations of words that capture the syntactic and semantic relationships.
Moreover, Chen et al. \cite{chen2020graph} covered various graph representation learning techniques, i.e., extracting the low-dimensional vector representations that preserve the intrinsic graph properties.
Graph representation learning has diverse applications, such as node classification \cite{bhagat2011node} and link prediction \cite{wang2015link}.
Due to the prevalence of representation learning in machine learning, in this paper, we mainly consider the sensitive information leakage risks during trading data representations.

\section{System Design}
In this section, we first mathematically formulate the information audit problem.
Then, in Section \ref{subsec: overview}, we show an overview of the FIAT system and briefly introduce each module of the design.
\subsection{Problem Formulation}
\label{subsec: problem_formulate}
To define the problem, it is important to identify the information that is legitimately restricted for transfer and the information that are allowed for free flows.
Regulated information stands for special categories of data that contain personal information that reveals a person's unique identity, including one's racial, political, and religious status, as specified in Art 9.1 of the European General Data Protection Regulation (GDPR) \cite{EUdataregulations2018}.
On the other hand, the data permitted for free flows are not strongly associated with personal identities, such as one's purchase history, click history, and economic status.
For notation convenience, we use \textit{sensitive data} to represent data that contain regulated information, and \textit{non-sensitive data} to represent other data that are allowed for free flows.

Denote the sensitive data as $\vec{\mathbf{x}}^{(s)} \in \mathcal{R}^n$ and the non-sensitive data as $\vec{\mathbf{x}}^{(ns)}\in \mathcal{R}^m$.
Here, $n$ and $m$ are the number of sensitive and non-sensitive features in the data, respectively.
Thus, a dataset can be separated into the sensitive part $X^{(s)}=\left\{\vec{\mathbf{x}}_i^{(s)}\right\}_{i=1}^N$ and the non-sensitive part $X^{(ns)}=\left\{\vec{\mathbf{x}}_i^{(ns)}\right\}_{i=1}^N$, where $N$ is the total number of data points.
The information auditing mainly focuses on revealing how much information about the sensitive data in $X^{(s)}$ is leaked by giving out the non-sensitive data in $X^{(ns)}$ or their representations.

As the raw data are also representations of themselves, we only discuss the information leakage in data representations for the rest of this paper.
During a specific data flow, a data receiver can request the representations of the data derived from a specific representation learning algorithm $f(\cdot)$ on the non-sensitive part of the data owned by a data sender.
Denote the representations of the data as $Y = \left\{\vec{\mathbf{y}}_i; \vec{\mathbf{y}}_i = f(\vec{\mathbf{x}}_i^{(ns)})\right\}_{i=1}^N$.
We measure the amount of leaked information by the mutual information $\mathcal{I}(X^{(s)};Y)$ between the sensitive data $X^{(s)}$ and data representations $Y$.
Additionally, an information leakage threshold $T$ is specified to limit the data receiver from exploiting the regulated information from data representations.
If the information leakage is above $T$, the data sender will reject to provide the required data representations.

\subsection{System Overview}
\label{subsec: overview}
Figure \ref{fig:system} shows an overview of the FIAT auditing system design.
Completing an information auditing requires 5 phases: Committing phase, Computation Proposing phase, Auditing phase, Verifying phase, and Decoding phase.
Below, we briefly describe the procedures of each phase for a data sender and a data receiver during a transborder data flow.
\subsubsection{Committing phase}
In order to prove the information leakage measurement on the data, the data sender will first need to \textit{commit} to the data to guarantee that the data used to generate the proof later will not be modified.
The easiest way to commit the data is to upload a hash value of the data onto the blockchain.
Here we require the data sender to hash both the sensitive and the non-sensitive features of the data because both parts will be used in auditing.
This hash value of the data can be easily stored on the blockchain, such as in smart contracts or in non-fungible tokens (NFTs).
Later in the \textit{Auditing Phase}, the data sender will only need to additionally prove that the committed data are used for auditing.

\subsubsection{Computation Proposing phase}
If a data receiver desires a specific kind of representation of the data, she can propose a representation learning algorithm $f(\cdot)$ on the data by uploading it to the smart contract, or simply stating she wants the raw data.
Additionally, the data receiver will need to upload a public key $pk$ for encryption to avoid the results getting public.
The smart contract will check the algorithm $f(\cdot)$ and public key $pk$ are supported before accepting the proposal.

\subsubsection{Auditing phase}
\label{subsubsec: audit}
During the computation auditing phase, the data sender performs the proposed computation on the non-sensitive data $X^{(ns)}$ and gets results $Y$.
To measure the amount of sensitive information leakage from the data representations $Y$, we propose to calculate the mutual information $\mathcal{I}(X^{(s)};Y)$ between the sensitive data $X^{(s)}$ and the representations $Y$.
The data sender checks whether the data representations can be given to the data receiver by comparing the mutual information with an information leakage threshold $T$.
If the leakage is above the threshold $T$, the data sender rejects the computation proposal by uploading $(reject, MI, \vec{\mathbf{0}})$ along with a zero-knowledge proof $\pi$ proving the computation of $MI$ is correct.
On the other hand, if the information leakage is acceptable, the data sender will encrypt the data representations with the data receiver's public key $pk$ and publish the encrypted results by uploading $(accept, MI, Y_{enc})$ along with a zero-knowledge proof $\pi$ proving that the computation of $Y$ and $MI$, and the encryption $Y_{enc}=Enc(pk, Y)$ are all done correctly.
In order to make the auditing process transparent to the data receiver, the data sender should upload a concrete threshold $T$ in \textit{Committing phase}, where $T$ can be estimated from the Shannon entropy \cite{shannonMathematicalTheoryCommunication2001} of the sensitive data.
The details for estimating the mutual information are illustrated in Section \ref{subsec: mi_estimate}, and the details of the zero-knowledge proof are described in Section \ref{subsec: proof_details}.

\subsubsection{Verifying phase}
After the data sender uploads the auditing results and the proof to the smart contract, the smart contract will automatically execute the verification process of the proof.
If the proof is valid, the auditing results and the required data representations will both be updated on the blockchain.
Otherwise, if the result is $reject$, the smart contract will simply record the calculated $MI$ as the reason for rejection, and the data receiver will not get the expected data representations.

\subsubsection{Decoding phase}
If the computation passes the auditing, the encrypted data representations will be stored in the smart contract.
Then, the data receiver can simply use her secret key $sk$ to decode the results and get $Y=Dec(sk, Y_{enc})$.
Moreover, as the whole auditing is through a zero-knowledge proof, the data receiver can be convinced that all the computations are carried out correctly.
Also, since the proof is zero-knowledge, the data receiver learns nothing about the data from the proof $\pi$, which cryptographically guarantees data privacy.
\section{Design Details}
In this section, we describe the design details of the FIAT system, including the security analysis, the selection of cryptographic algorithms, the information intensity estimation of sensitive features, the details in ZKP and the smart-contract-based risk management system.

\subsection{System Security}
In FIAT system, we consider the following 4 requirements for the security of the system:
\begin{enumerate}
    \item \textit{Correctness:} If the data sender is honest, she should be able to convince the data receiver that the auditing is carried out correctly.
    \item \textit{Soundness:} A malicious data sender cannot use different data for auditing, report wrong mutual information results, or upload incorrect encrypted data representations.
    \item \textit{Privacy-preserving:} The data receiver cannot get any information on the data from the auditing results and the proof. Moreover, the data receiver cannot get any information about the data representations if the auditing fails.
    \item \textit{Liveness:} The data receiver can finally know the auditing results, and can get the requested data representations if the auditing passes. 
\end{enumerate}

The first three properties are guaranteed by the securities of the underlying hash function, the zero-knowledge proof protocol, and the public-key encryption scheme.
For liveness, there is a certain possibility that the data sender rejects to audit the computation after the data receiver uploads the computation proposal.
However, such unresponsiveness can usually lead to negative social or economical impacts on the data sender.
Therefore, the liveness of FIAT system is reinforced through social, governmental, and economical relationship such as legitimate contracts and laws.
\subsection{Hash Function \& Encryption Scheme}
To optimize the ZKP performance, it is crucial to choose protocols that are zk-friendly to minimize the proof generation and verification time.
Therefore, we choose \textit{Poseidon hash} as the hash function and \textit{ECIES} with \textit{MiMC cipher} as the public-key encryption scheme.

\subsubsection{Poseidon Hash}
Poseidon \cite{grassiPoseidonNewHash2021} is a hash function designed to minimize the number of constraints in R1CS (Rank-1 Constraint System), one kind of arithmetic circuit representation widely used in current zk-SNARK protocols.
Fewer constraints imply smaller circuit size and thus faster computation for proof generation.
Moreover, Poseidon also works naturally in a prime field, which perfectly fits zk-SNARK circuits.

\subsubsection{Elliptic Curve Integrated Encryption Scheme (ECIES)}
We choose ECIES as the asymmetric encryption scheme for the convenience of implementation in the prime field.
In ECIES, the data sender and the data receiver first perform an elliptic-curve-based key exchange (a.k.a. Elliptic Curve Diffie–Hellman \cite{bernstein2006curve25519}) to establish a shared symmetric encryption key.
Then, they use the encryption key to perform symmetric encryption and decryption over the secret data.
For the symmetric encryption, we choose \textit{MiMC cipher} \cite{albrechtMiMCEfficientEncryption2016} as the encryption scheme to optimize for arithmetic circuits.
MiMC encryption uses a very small number of multiplicative operations and is operated in prime fields, which makes it a suitable block cipher for zk-SNARK circuits.

\subsection{Information Intensity Estimation of Sensitive Features}
\label{subsec: mi_estimate}
In Section \ref{subsubsec: audit}, we mentioned that the computation of mutual information $\mathcal{I}(X^{(s)};Y)$ between the results $Y$ and the sensitive data $X^{(s)}$ is conducted in the auditing phase of the FIAT system.

Assume all the data points are independently sampled from the same distribution, i.e., the samples are \textit{iid}.
By the definition of mutual information, we have
\begin{equation}
    \mathcal{I}(X^{(s)};Y)=\int_{\A} \int_{\B} P_{X^{(s)}Y}(\A,\B)\ln\frac{P_{X^{(s)}Y}(\A,\B)}{P_{X^{(s)}}(\A)P_{Y}(\B)}\text{d}\A\text{d}\B,
\end{equation}
where $X^{(s)}$ and $Y$ are random variables defined above, and $P_{X^{(S)}}(\cdot)$, $P_{Y}(\cdot)$, and $P_{X^{(S)}Y}(\cdot,\cdot)$ denote their respective distributions and the joint distribution.
In most real-world cases, we may not know the real distributions.
Therefore, we need to estimate them from the data points.

We adopt the kernel density estimation (KDE) to estimate the distribution $P_{X^{(s)}}(\cdot)$.
KDE is a non-parametric estimation for fitting the distribution of a random variable.
We estimate $P_{X^{(s)}}(\cdot)$ by
\begin{equation}
    \hat{P}_{X^{(s)}}(\cdot)=\frac{1}{N}\sum_{i=1}^{N} K_{\D}(\cdot, \vec{\mathbf{x}}_i^{(s)}),
\end{equation}
where $N$ is the size of the data, $\vec{\mathbf{x}}_i^{(s)}$ is the $i^{th}$ data point in $X^{(s)}$, $\D$ is a parameter called \textit{bandwidth}.
$K_{\D}(\vec{\mathbf{x}}_1,\vec{\mathbf{x}}_2)$ is the kernel function defined as 
$\mathbbm{1}\left\{\left\lVert\left\lfloor\frac{\vec{\mathbf{x}}_1}{\D}\right\rfloor\D-\vec{\mathbf{x}}_2\right\rVert\leq \D\right\}$ which equals to $1$ when $\left\lVert\left\lfloor\frac{\vec{\mathbf{x}}_1}{\D}\right\rfloor\D-\vec{\mathbf{x}}_2\right\rVert\leq \D$ and $0$ otherwise.
$\left\lfloor\frac{\vec{\mathbf{x}}_1}{\D}\right\rfloor$ is to take the largest integer below $\frac{\vec{\mathbf{x}}_1}{\D}$ for each dimension.
We can estimate $P_{Y}(\cdot)$ and $P_{X^{(s)}Y}(\cdot,\cdot)$ in a similar way.
Note that as $\D$ converges to $\vec{\mathbf{0}}$, the estimated distributions $\hat{P}_{X^{(s)}}(\cdot)$, $\hat{P}_{Y}(\cdot)$, $\hat{P}_{X^{(s)}Y}(\cdot,\cdot)$ converge to the real distributions $P_{X^{(s)}}(\cdot)$, $P_{Y}(\cdot)$, $P_{X^{(s)}Y}(\cdot,\cdot)$.
Therefore, we deduce an unbiased estimation of the mutual information using the estimated distributions
\begin{equation}
    \hat{\mathcal{I}}(X^{(s)};Y)=\int_{\A} \int_{\B} \hat{P}_{X^{(s)}Y}(\A,\B)\ln\frac{\hat{P}_{X^{(s)}Y}(\A,\B)}{\hat{P}_{X^{(s)}}(\A)\hat{P}_{Y}(\B)}\text{d}\A\text{d}\B.
\end{equation}
In the above calculation, the values of $\hat{P}_{X^{(s)}}$, $\hat{P}_{Y}$, and $\hat{P}_{X^{(s)}Y}$ are constant on each hypercube of the form $X_C=\bigotimes_{i}[\S_i \odot \D_{X},\S_{i+1}\odot\D_{X}]$, $Y_C=\bigotimes_{i}[\T_i\odot\D_{Y},\T_{i+1}\odot\D_{Y}]$, and $X_C\bigotimes Y_C$ respectively, where $\S_i$ and $\T_i$ are integer vectors indicating the index of each hypercube, $\odot$ denotes the element-wise product, and $\bigotimes$ denotes the union of intervals.
Thus, in the practical computation, the integral can be reduced to the summation over all hypercubes, which becomes
\begin{equation}
    \hat{\mathcal{I}}(X^{(s)};Y)=\sum_{\A\in X_C}\sum_{\B\in Y_C} \hat{P}_{X^{(s)}Y}(\A,\B)\ln\frac{\hat{P}_{X^{(s)}Y}(\A,\B)}{\hat{P}_{X^{(s)}}(\A)\hat{P}_{Y}(\B)}.
\end{equation}

As also shown in Section \ref{subsec: problem_formulate}, we define $Y$ as the set of results computed only on single data points, resulting in one data representation per data point.
This definition simplifies the calculation of mutual information in zero-knowledge proofs.
We leave the information auditing for more general computing functions as future work.

\begin{algorithm}

\caption{Computation Auditing}
\label{algo: proof_generate}
\begin{algorithmic}
\Procedure{Auditing}{$X^{(s)}$, $X^{(ns)}$, $f$, $pk$, $H$, $T$}
\State Generate a zero-knowledge proof $\pi$ for the following computations:

\State $H \leftarrow hash(X^{(s)}, X^{(ns)})$ \Comment{Hash Check} 
\State $Y \leftarrow f(X^{(ns)})$
\State $MI \leftarrow \mathcal{I}(X^{(s)};Y)$ \Comment{Auditing}

\If{$MI > T$}
    \State $pass ~\leftarrow~ reject$
    \State $Y_{enc}\leftarrow \vec{\mathbf{0}}$
\Else 
    \State $pass ~\leftarrow~ accept$
    \State $Y_{enc} \leftarrow Enc(pk, Y)$  \Comment{Encryption}
\EndIf\\
\quad\ \ \Return $(pass, MI, Y_{enc}, \pi)$
\EndProcedure
\end{algorithmic}
\end{algorithm}

\subsection{Proof for Correct Auditing}
\label{subsec: proof_details}
Protocol \ref{algo: proof_generate} shows the whole auditing process of the data sender.
The main task of the data sender is to generate a zero-knowledge proof to prove that the computation is carried out correctly, without leaking any information about the secret data.
Afterwards, the proof can be publicly verified through the smart contract. 

Totally four calculations should be included in the proof to pass the verification.
The first calculation is to prove the data used for the computation auditing are the same as the previously committed one.
The data sender should prove that the hash value of the data equals to the one stored in the smart contract.
The collision-resistance property of the hash function guarantees that it is impossible for the data sender to generate different data with the same hash value as the committed data.
The second calculation is to prove the data representations are correctly derived from the raw data.
Although the representations $Y$ are not publicly revealed, they are taken as the inputs to the calculations afterwards.
Thus, the correctness of the calculations afterwards can imply the correctness of this calculation.
The third calculation is to prove the correctness of the mutual information calculation $\mathcal{I}(X^{(s)};Y)$.
As the function to calculate mutual information is pre-specified, the data sender cannot make wrong proofs for this calculation.
Finally, in order to protect the privacy of the data representations and make them only viewable to the data receiver, the data sender needs to encrypt the data representations with the public key $pk$ and the pre-specified encryption scheme.
The encryption process also needs to be proved to convince the data receiver.
These four computations can be proved in one single zero-knowledge proof and verified on the smart contract.
Afterwards, the data sender only needs to decide whether to share the encrypted data representations $Y_{enc}$ to the data receiver, depending on whether the information leakage is above or below the threshold $T$.

The design of the FIAT system does not require specifying which zero-knowledge proof protocol to use.
Yet, different protocols have different trade-offs on security and efficiency and should be picked carefully for specific use cases.

\begin{algorithm}

\caption{Auditing Contract}
\label{algo: verify_contract}
\begin{algorithmic}
\State $MI_{total}:=0$
\Procedure{GetData}{$H$, $T$, $\mathcal{I}$, $Enc$}
\State require($msg.sender == owner \And DataHash == \perp$) 
\State  $DataHash \leftarrow H$
\State  $MI_{threshold} \leftarrow T$
\State  $AuditFunc \leftarrow \mathcal{I}$ 
\State  $EncFunc \leftarrow Enc$
\EndProcedure
\Procedure{GetProposal}{$f$, $pk$}
\State require($msg.sender == consumer$)
\State $Algo \leftarrow f$
\State  $Pubkey \leftarrow pk$
\EndProcedure
\Procedure{VerifyAndUpdate}{$pass$, $MI$, $Y_{enc}$, $\pi$}
\State require($msg.sender == owner$)
\If{$pass == accept \And MI \leq MI_{threshold}$}
\State Verify $\pi$ is a valid proof. If not, return $\perp$.
\State $MI_{total} \leftarrow MI$
\State $Result \leftarrow Y_{enc}$
\ElsIf{$pass == reject \And MI > MI_{threshold}$}
\State Verify $\pi$ is a valid proof. If not, return $\perp$.
\State $MI_{total} \leftarrow MI$
\State $Result \leftarrow \vec{\mathbf{0}}$ \Comment{$Y_{enc}=\vec{\mathbf{0}}$}
\EndIf
\EndProcedure
\end{algorithmic}
\end{algorithm}

\subsection{Smart-Contract-based Risk Management}
Along with the auditing, we embed a simple risk management system in the smart contract for proof verification.
Since the smart contract is a good place to store data that require consensus, we use the verification contract to record the auditing results.
Meanwhile, the smart contract also checks if the final decision of the data sender is valid, i.e., the data representations should not be published if the information leakage is above the threshold $T$.
The smart contract will only accept the auditing results if both the zero-knowledge proof and the auditing decision are verified.
The complete design of the auditing contract is shown in Protocol \ref{algo: verify_contract}.
\section{Experiments}
We conduct extensive experiments to visualize the information leakage risk and measure the performance of the FIAT auditing system.
In this section, we introduce the dataset and the data representation learning algorithm used in our experiments in Section \ref{subsec: dataset}, present the results of visualizing the information leakage risk in Section \ref{subsec: quantify}, and show the performance of the FIAT system in Section \ref{subsec: zkp_exp}.

\subsection{Dataset \& Methods}
\label{subsec: dataset}
For all the experiments, we use the Adult Data Set \cite{Dua:2019} as the data used in the transborder data flow.
Adult Data Set is a public dataset containing personal information collected from different countries, originally used to predict each person's income based on their profiles.
The personal information contains \textit{age}, \textit{workclass}, \textit{education}, \textit{education-num}, \textit{marital-status}, \textit{occupation}, \textit{relationship}, \textit{race}, \textit{gender}, \textit{capital-gain}, \textit{capital-loss}, \textit{hours-per-week}, and \textit{native-country}.
Among all the features, we take \textit{age}, \textit{marital-status}, \textit{relationship}, and \textit{race} as the regulated sensitive features because those are usually private information that strongly relates to a person's unique identity.
Other features are all considered non-sensitive.
For each categorical feature, we project the categories into indexes and use the indexes as the raw inputs for all the following experiments.

We choose \textit{Principal Component Analysis (PCA)} \cite{pearsonLIIILinesPlanes1901} as the data representation learning algorithm for the demonstration of the information leakage risks and the evaluation of the FIAT system.
PCA is a classic representation learning algorithm that reduces high-dimensional data to low-dimensional data while preserving the most significant information in the original data.
Many machine learning tasks adopt PCA in the data preprocessing phase to improve the training efficiency while remaining the original data characteristics.

\subsection{Visualizing the Information Auditing Results with Machine Learning}
\label{subsec: quantify}
\begin{figure}[htbp]
    \centering
    \includegraphics[width=0.85\columnwidth]{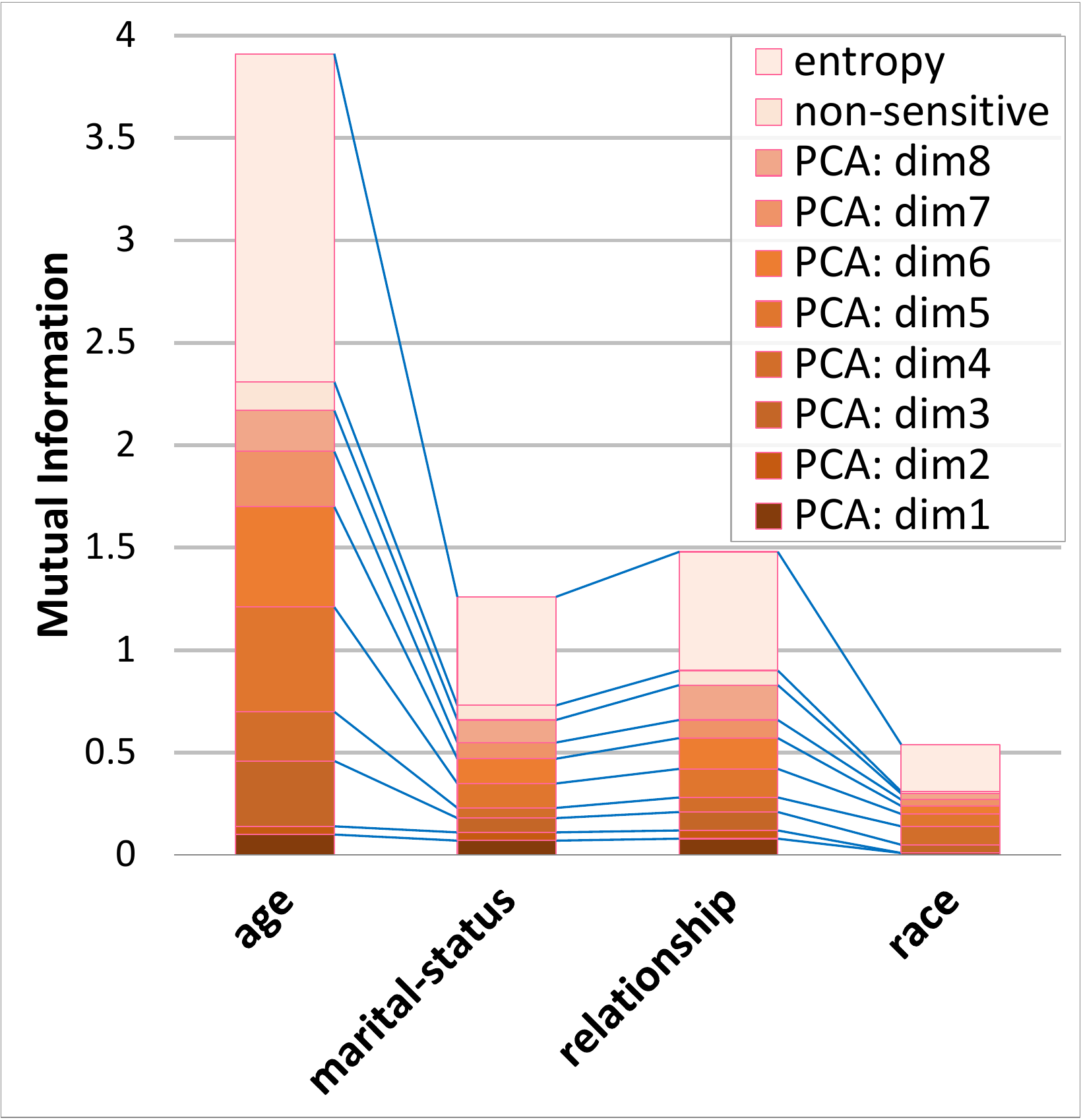}
    \caption{Information leakage of each sensitive feature from data representations obtained from PCA algorithm (and the raw data), compared with their Shannon entropy.}
    \label{fig:result_mi}
\end{figure}
\subsubsection{Experimental Setup}
In this experiment, we measure the mutual information between each sensitive feature and the data representations obtained through the PCA algorithm.
Since the original non-sensitive data have $9$ dimensions, we use PCA to reduce these data to $1\sim8$ dimension(s) separately and measure the sensitive information leakage for each PCA output.
Additionally, we calculate the mutual information between sensitive features and all the other non-sensitive features as the amount of information leakage by directly giving out the raw data.
We compare the amount of information leakage with the maximum information of each sensitive feature, measured by Shannon entropy \cite{shannonMathematicalTheoryCommunication2001}.

Moreover, to reveal the risk of information leakage, i.e., how the data receiver can predict the regulated information using the obtained data representations, we train a simple Multi-layer Perceptron (MLP) to predict each sensitive feature from the data representations, which gives a lower bound on the regulated information leakage risk.
The MLP has only one hidden layer with $64$ nodes and ReLU \cite{nair2010rectified} activation.
We train each prediction task for $200$ epochs using Cross-entropy Loss and Adam \cite{kingma2014adam} optimizer with $10^{-3}$ initial learning rate.
Especially, since \textit{age} is the only continuous feature, we divide it into intervals with length $10$ and treat the intervals as different classes for classification.
We randomly take $90\%$ of the dataset for training and $10\%$ for testing and report the average test accuracy of $5$ randomly-initiated runs minus the Zero Rate Classifier (ZeroR) accuracy, i.e., trivially predict the largest class, as the practical accuracy gain from knowing the data representations.

\begin{figure}[htbp]
    \centering
    \includegraphics[width=0.9\columnwidth]{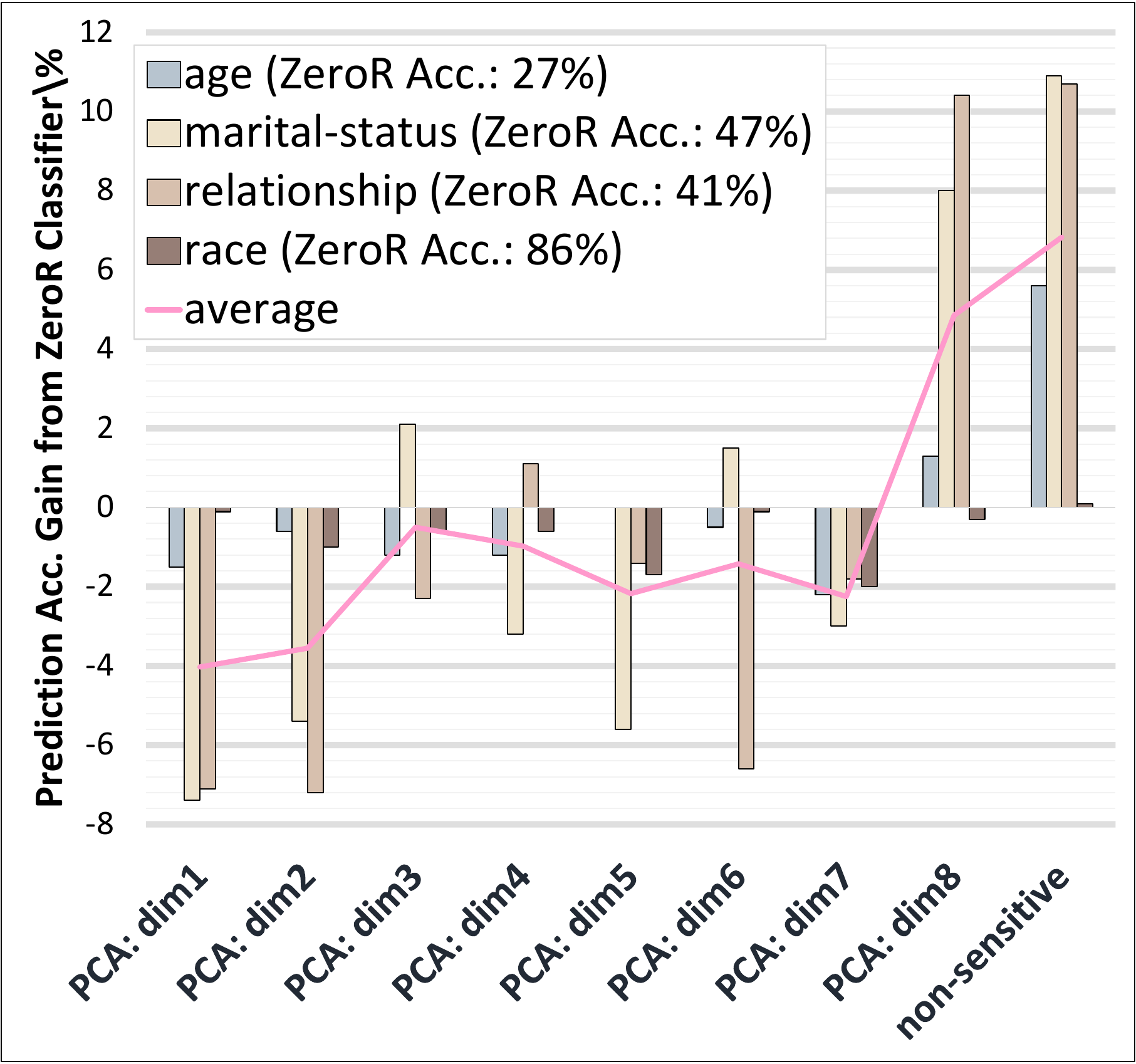}
    \caption{Test accuracy gain of predicting each sensitive feature using the data representations obtained from PCA algorithm (and the raw data) compared with Zero Rate Classifier (ZeroR). The pink line shows the average accuracy gain of $4$ sensitive features for each data representation.}
    \label{fig:result_acc}
\end{figure}
\subsubsection{Results}
Figure \ref{fig:result_mi} shows the information leakage of each sensitive feature measured by the mutual information with the data representations obtained from the PCA algorithm.
We also add the mutual information between the sensitive features and the raw data, and the Shannon entropy of each sensitive feature for comparison.
It is clear that the non-sensitive features contain more than half of the information of each sensitive feature.
With the decrease of the dimension of data representations, the information leakage gradually reduces to a very small percentage.
The decrease is more obvious in features with large entropy than those with small entropy.
Moreover, when the dimension is reduced below $4$, the information leakage drops to almost $0$ for all the sensitive features, which informs that it is feasible to reduce the amount of sensitive information leakage while preserving other important information for machine learning.

Figure \ref{fig:result_acc} presents the prediction accuracy for each sensitive feature with a simple 3-Layer MLP as the model and the data representations as the inputs.
The figure shows the test set accuracy minus the Zero Rate Classifier accuracy, which represents the practical accuracy gained from knowing the data representations.
These statistics show a lower bound for the sensitive information leakage risks, since stronger machine learning models can get better prediction results.
It can be observed that all the data representations with less than $8$ dimensions have negative accuracy gains over almost all the sensitive features.
Conversely, for PCA representations with $8$ dimensions and the non-sensitive raw data, the average accuracy gain is around $5\sim 7\%$, with the largest gain about $11\%$.
As a result, directly transferring the non-sensitive data or the data that are not sufficiently processed can both lead to potential exploitation of the regulated information, which shows the importance of information auditing in data flows.
Combining Figure \ref{fig:result_mi} and \ref{fig:result_acc}, it is safe to give out the data representations when the mutual information is below $40\%$ of the original entropy, which serves as a reference to set the information leakage threshold $T$ in the FIAT system.
\subsection{Benchmarking the FIAT System}
\label{subsec: zkp_exp}
\subsubsection{Experimental Setup}
We use \textbf{circom} \cite{Circom2022} to implement the auditing circuit and \textbf{snarkjs} \cite{Snarkjs2022} to generate witness, proof and the verification contract.
\textbf{circom} is an open-source programming language specially designed for writing and compiling ZKP circuits to R1CS constraints.
\textbf{snarkjs} is a commonly used zk-SNARK library written in JavaScript for proof generation and verification, which is compatible with the R1CS representations generated by \textbf{circom}.

The entire auditing circuit consists of four fundamental components, i.e., the \textit{Poseidon Hash} commitment, the \textit{ECIES} encryption, the \textit{PCA} calculation, and the \textit{Mutual Information (MI)} calculation.
The \textit{ECIES} encryption consists of an implementation of an \textit{ECDH} key exchange scheme and a \textit{MiMC Cipher} for symmetric encryption.
We directly leverage the implementation of \textit{Poseidon Hash} and \textit{MiMC Cipher} from \textbf{circomlib} \cite{CircomLib2022} which is also an open-source library that implements several basic circuits for cryptography in \textbf{circom}.
The circuits for the PCA algorithm, the MI computation, and the ECDH key exchange are implemented from scratch.
The details of these implementations are illustrated in Appendix \ref{appdix: implementation}.

To thoroughly test the performance of our implementation, we conduct extensive benchmarking tests on the auditing circuit in the FIAT system.
Firstly, we measure the number of R1CS constraints generated from the auditing circuit for datasets with different sizes, which is a critical metric to measure the circuit size for the estimation of ZKP performance.
The size of the dataset ranges continuously from $500$ to $5000$ with the interval $500$.
To identify the computation overhead of the whole system, we separately benchmark the above-mentioned four modules of the auditing circuit and plot their relation with dataset size in Figure \ref{fig:num_constraint}.

In addition to the study of the number of constraints, it is also important to quantify the computational resource consumption for generating and verifying the proof.
To this end, we adopt the Groth16 \cite{grothSizePairingBasedNoninteractive2016} algorithm implemented in \textbf{snarkjs} to generate and verify the proof of our auditing circuit.
We report the running time of each step in proof generation and verification with dataset size varying from $500$ to $1500$, including the circuit compilation time, the witness generation time, the proving time, and the verifying time.
We also present the proving key size and the proof size to identify the storage requirements.
The benchmark is conducted on an \texttt{AWS EC2} server with $128$ CPUs and $496$ GB RAM.
All the statistics are reported in Table \ref{tab:proof_time}.
To save the setup time, we download the pre-generated power of tau file  \cite{PoT} provided by \textbf{snarkjs}.

\subsubsection{Results}
\begin{figure}[htbp]
    \centering
    \includegraphics[width=0.95\columnwidth]{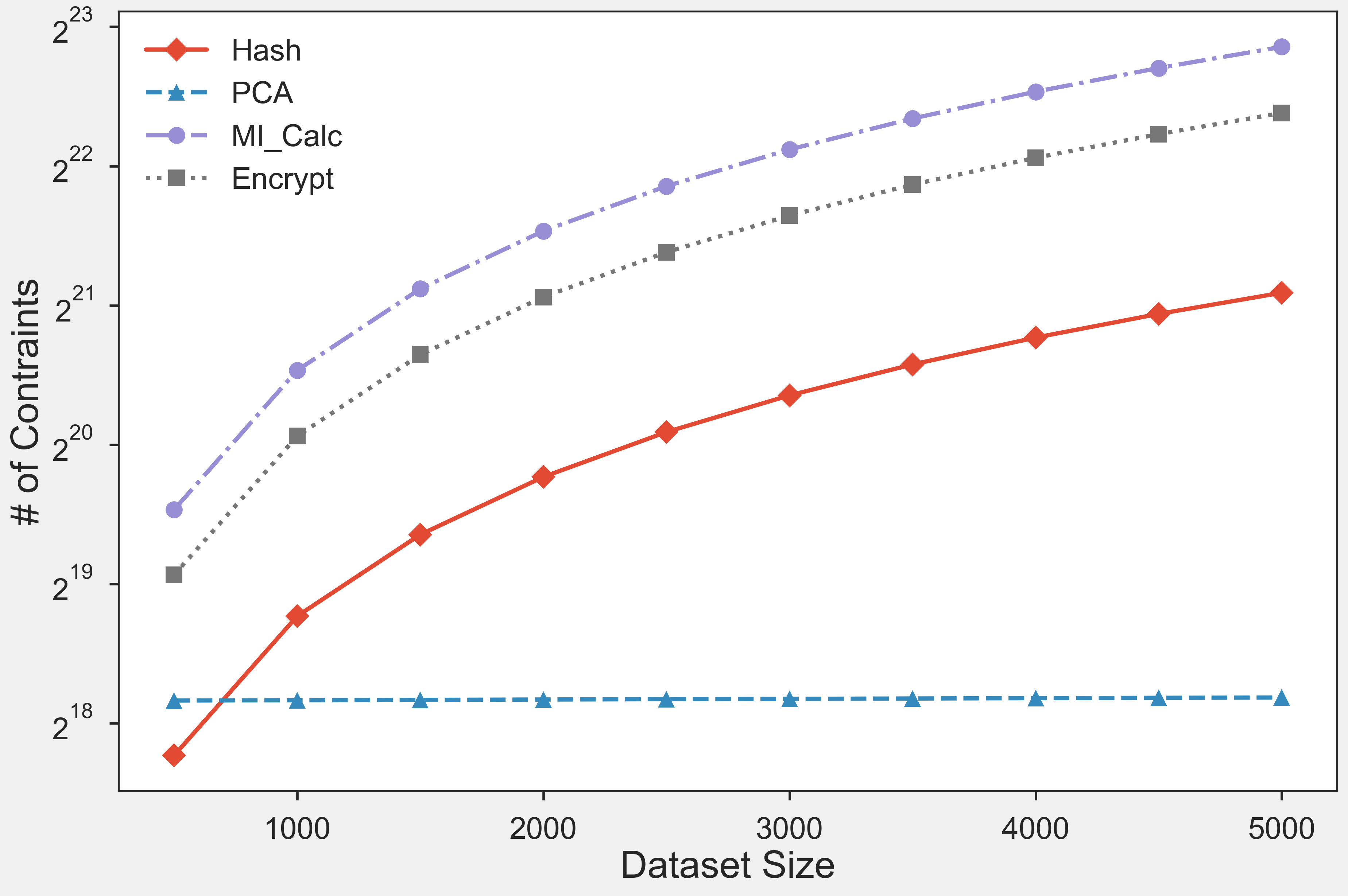}
    \caption{Relation between dataset size and \# of R1CS constraints for each part of computation in zero-knowledge proof circuits.}
    \label{fig:num_constraint}
\end{figure}

As shown in Figure \ref{fig:num_constraint}, the number of R1CS constraints grows linearly with the dataset size, with $<2^{24}$ constraints for a dataset with $5000$ person information entries.
It is shown that using advanced zk-SNARK protocols, such circuits can be proved within several minutes \cite{drevonBenchmarkingZeroKnowledgeProofs}, which perfectly satisfies the demand for data auditing in the data market.
For the comparison of each computation module, the encryption and the mutual information calculation take up most part of the auditing circuit size.
This is because the encryption is done point-by-point, and the mutual information computation involves the computation of the kernel function, which is relatively heavy to be implemented in circuits.
The PCA part generates only a few and almost constant constraints because the size of the covariance matrix is very small and irrelevant to the size of the dataset.

For the benchmarking part of the proof generation and verification, as shown in Table \ref{tab:proof_time}, it is clear that the trusted setup step in Groth16 takes up most of the time.
However, as the compilation and the setup step are once-for-all for the same circuit, once they are completed, all the proof generation in the future will be much faster, only lower bounded by the witness generation time and the proving time, which is about $2-3$ minutes in total.
A great advantage of Groth16 is that the proof size and verification time are both small and constant for all the circuits, which makes our auditing system suitable to be validated in smart contracts and scalable for large datasets.
Moreover, the storage required to generate the proof is within $GB$ scale, which is not a big problem for modern computation clusters.
\begin{table}
    \centering
    \begin{tabular}{lrrr}
        \toprule
        \textbf{Dataset Size}& $\mathbf{500}$ & $\mathbf{1000}$ & $\mathbf{1500}$\\
        \midrule
        \# of R1CS constraints & $1683419$& $3288907$ & $4894395$\\
        Circuit compilation & $140s$ & $219s$ & $300s$\\
        Witness generation & $1s$ & $1s$ & $2s$\\
        Setup time & $599s$ & $1023s$ & $1965s$\\
        Proving key size & $1.2GB$ & $2.3GB$ & $3.5GB$\\
        Proving time & $47s$ & $89s$ & $137s$\\
        Proof size & $0.8KB$ & $0.8KB$ & $0.8KB$\\
        Proving verification time & $2s$ & $2s$ & $2s$ \\
        \bottomrule
    \end{tabular}
    \caption{Benchmarking statistics of ZKP generation and verification using Groth16 in the FIAT system for dataset with different size.}
    \label{tab:proof_time}
\end{table}
\section{Conclusion}
In this paper, we emphasize the necessity of auditing the regulated information leakage for transborder data flows, an important problem that has long been neglected by previous research on data privacy.
We propose to measure such information leakage risks using mutual information and design a designated auditing system \textbf{FIAT} to achieve \textbf{F}ine-grained \textbf{I}nformation \textbf{A}udit for \textbf{T}rustless transborder data flows.
Specially, we leverage the power of zero-knowledge proof to protect data privacy during auditing and adopt smart contracts to make the auditing results publicly verifiable and trustworthy.
To unveil the information leakage risks concretely, we conduct experiments to show the relation between the amount of information leakage and the resulting predictability on a real-world dataset.
We also conduct benchmark the performance of the FIAT system in terms of the zero-knowledge proof generation and verification.
Results show that the FIAT system can complete a large amount of data auditing in a small period of time, making it practical to be adopted for real-world data flows.

\appendix

\section{Details in Auditing Circuit Implementation}
\label{appdix: implementation}
In this section, we present some implementation details of our ZKP auditing circuit for better understanding.
\subsection{Fast Matrix Multiplication Verification}
A normal matrix multiplication for two matrices $A$ and $B$ with $m\times n$ and $n\times k$ dimensions requires $O(mnk)$ element multiplications.
However, verifying that $A \cdot B = C$ has a faster algorithm with randomness, which means the multiplication computation can be conducted off-circuit and verified in the circuit.
The trick is that we generate a random vector $r$ with $m$ dimensions, and multiply both sides of the equation by $r$ to get
\begin{equation}
    A\cdot Br = Cr.
    \label{eq: matmul}
\end{equation}
Computing Eq. \ref{eq: matmul} only requires $O(mn)$ element multiplications.
Repeating the verification for several different $r$ can reduce the soundness error of such verification to a very small level.
\subsection{Floating Point \& Fixed Point}
The computation of PCA and mutual information require floating-point representations, while the ZKP circuit can only operate fixed-point numbers in a prime field.
To resolve such conflicts, we use a similar strategy as Mystique \cite{wengMystiqueEfficientConversions2021} to project the floating-point values to the prime field with only a small loss of precision.
Specifically, each floating-point number is first multiplied with $2^s$ where $s$ is the precision we want.
In our system, we take $s=20$.
Then, each number within a representable range is projected to the interval $[0, p-1]$ and normally operated as fixed-point numbers, where $p$ is the prime number for the prime field.
The addition and multiplication operations are the same for the floating-point numbers and the fixed-point numbers.
\subsection{PCA Circuit}
The PCA algorithm consists of some matrix additions and multiplications, and a Singular Value Decomposition (SVD) for extracting eigenvectors.
For the SVD part, we implement the Power Iteration algorithm to extract eigenvalues and eigenvectors one-by-one from the covariance matrix.
For the Power Iteration algorithm, since it is infeasible to check convergence within circuits before knowing the inputs, we simply set a maximum of $20$ iterations for each extraction.
For all the proof generation and verification benchmarking, we use PCA to reduce the original data to $3$ dimensions.

\subsection{Mutual Information Circuit}
The heaviest part of computation in the mutual information circuit is the kernel function computation, in which each number needs to be classified to a specific interval for further computation.
For each number in PCA results, we classify them into $10$ intervals.
Later, we simply count the numbers in the intervals to calculate the probability and the mutual information, although this part generates a large number of constraints because the values are not known in advance during circuit compilation.
For the logarithm computation, we also put it off-circuit and verify it in the circuit, simply with exponentiation.

\section{Formal Definition of Zero-knowledge Proof for $\mathcal{NP}$ Relations}
\label{appendix:zkp_def}
In this section, we rigorously present the mathematical definition of zero-knowledge proof for $\mathcal{NP}$ relations in Definition \ref{def: zkp_np}. 
\begin{definition}[Zero-knowledge Proof for $\mathcal{NP}$]
Denote $\lambda$ as the security parameter and $negl(\lambda)$ as a negligible number w.r.t. $\lambda$.
Let $(x, w)\in R$ be an instance of the relation $R$ specified by an $\mathcal{NP}$ language $L$, where $x$ is called the \textit{statement} and $w$ is called the \textit{witness}.
$x \in L$ if and only if there exists $w$ such that $(x, w)\in R$.
We say a pair of interactive Turing machines $\langle P,V\rangle$ is a zero-knowledge interactive proof system for $R$ if the following three properties are satisfied:
\begin{enumerate}
    \item \textbf{Completeness:} For any $(x,w) \in R$ and an honest prover $P$, the verifier $V$ accepts the proof with probability at least $ 1-negl(\lambda)$.
    \item \textbf{Knowledge Soundness:} 
    For any malicious prover $P^*$ that makes $V$ accept, there exists a Probabilistic Polynomial Time (PPT) extractor $\mathcal{X_{P^*}}$ with full access to $P^*$'s state such that
    \begin{equation}
    \begin{aligned}
        &\Pr\{Accept \leftarrow \langle P^*(\lambda, x),V(\lambda, x)\rangle, \\
        &\quad \quad \quad w \leftarrow \mathcal{X_{P^*}}(\lambda, x): (x, w)\notin R\} \leq negl(\lambda).
    \end{aligned}
    \end{equation}
    \item \textbf{Zero Knowledge:} For any $(x, w)\in R$, there is a PPT simulator $Sim$ such that for any PPT distinguisher $\mathcal{A}$
    \begin{equation}
    \begin{aligned}
    &\big|\Pr\{\pi \leftarrow \langle P(
    \lambda, x, w),V(\lambda, x)\rangle: \mathcal{A}(\lambda, \pi)=1\} - \\
    &\quad\Pr\{\pi^* \leftarrow Sim(\lambda, x): \mathcal{A}(\lambda, \pi^*)=1\}\big| \leq negl(\lambda).
    \end{aligned}
    \end{equation}
\end{enumerate}
\label{def: zkp_np}
\end{definition}
In the circuit satisfiability problem, the circuit structure is the statement and the inputs of the circuit are the witness.

\bibliographystyle{unsrtnat}
\bibliography{bib}

\end{document}